\documentclass[10pt, conference]{IEEEtran}
\IEEEoverridecommandlockouts
\usepackage{ifdraft}

\usepackage[T1]{fontenc}
\usepackage[utf8]{inputenc}
\usepackage{amsmath}
\usepackage{mathrsfs}
\usepackage{hyperref}
\usepackage{wasysym}
\usepackage{yfonts}
\usepackage[plain]{fancyref}
\usepackage{subcaption}

\usepackage[final]{listings}
\usepackage{booktabs}
\usepackage{multirow}
\usepackage{xcolor,colortbl}
\usepackage{xspace}
\usepackage{suffix}
\usepackage{etoolbox}
\usepackage[inline]{enumitem}
\usepackage{acronym}
\usepackage{url}

\usepackage[backend=bibtex,
   style=ieee,
   citestyle=numeric-comp,
   natbib=true,
   maxnames=7,
   minnames=1,
   maxcitenames=1, 
   mincitenames=1,
   giveninits=true,
   hyperref=true,
   sorting=none,
   defernumbers]{biblatex}

\def\fref{\Fref} % treat all \frefs as \Frefs
\renewcommand{\lstlistingname}{Snippet}
\newcommand*{\fancyreflstlabelprefix}{lst} % define lst delimiter
\newcommand*{\Freflstname}{\lstlistingname}
\newcommand*{\freflstname}{\lstlistingname}
\Frefformat{vario}{\fancyreflstlabelprefix}%
  {\Freflstname\fancyrefdefaultspacing#1#3}
\frefformat{vario}{\fancyreflstlabelprefix}%
  {\freflstname\fancyrefdefaultspacing#1#3}
\Frefformat{plain}{\fancyreflstlabelprefix}%
  {\Freflstname\fancyrefdefaultspacing#1}
\frefformat{plain}{\fancyreflstlabelprefix}%
  {\freflstname\fancyrefdefaultspacing#1}

\newcommand*{\fancyrefthmlabelprefix}{thm} % define thm delimiter
\newcommand*{\Frefthmname}{Theorem}%
\newcommand*{\frefthmname}{%
 \MakeLowercase{\Frefthmname}}%
\Frefformat{vario}{\fancyrefthmlabelprefix}%
  {\Frefthmname\fancyrefdefaultspacing#1#3}
\frefformat{vario}{\fancyrefthmlabelprefix}%
  {\frefthmname\fancyrefdefaultspacing#1#3}
\Frefformat{plain}{\fancyrefthmlabelprefix}%
  {\Frefthmname\fancyrefdefaultspacing#1}
\frefformat{plain}{\fancyrefthmlabelprefix}%
  {\frefthmname\fancyrefdefaultspacing#1}

\newcommand*{\fancyreflemlabelprefix}{lem} % define lem delimiter
\newcommand*{\Freflemname}{Lemma}%
\newcommand*{\freflemname}{%
 \MakeLowercase{\Freflemname}}%
\Frefformat{vario}{\fancyreflemlabelprefix}%
  {\Freflemname\fancyrefdefaultspacing#1#3}
\frefformat{vario}{\fancyreflemlabelprefix}%
  {\freflemname\fancyrefdefaultspacing#1#3}
\Frefformat{plain}{\fancyreflemlabelprefix}%
  {\Freflemname\fancyrefdefaultspacing#1}
\frefformat{plain}{\fancyreflemlabelprefix}%
  {\freflemname\fancyrefdefaultspacing#1}

\newcommand*{\fancyrefdeflabelprefix}{def} % define def delimiter
\newcommand*{\Frefdefname}{Definition}%
\newcommand*{\frefdefname}{%
 \MakeLowercase{\Frefdefname}}%
\Frefformat{vario}{\fancyrefdeflabelprefix}%
  {\Frefdefname\fancyrefdefaultspacing#1#3}
\frefformat{vario}{\fancyrefdeflabelprefix}%
  {\frefdefname\fancyrefdefaultspacing#1#3}
\Frefformat{plain}{\fancyrefdeflabelprefix}%
  {\Frefdefname\fancyrefdefaultspacing#1}
\frefformat{plain}{\fancyrefdeflabelprefix}%
  {\frefdefname\fancyrefdefaultspacing#1}

\newcommand*{\fancyreflnlabelprefix}{ln} % define ln delimiter
\newcommand*{\Freflnname}{Line}%
\newcommand*{\freflnname}{%
 \MakeLowercase{\Freflnname}}%
\Frefformat{vario}{\fancyreflnlabelprefix}%
  {\Freflnname\fancyrefdefaultspacing#1#3}
\frefformat{vario}{\fancyreflnlabelprefix}%
  {\freflnname\fancyrefdefaultspacing#1#3}
\Frefformat{plain}{\fancyreflnlabelprefix}%
  {\Freflnname\fancyrefdefaultspacing#1}
\frefformat{plain}{\fancyreflnlabelprefix}%
  {\freflnname\fancyrefdefaultspacing#1}

\lstdefinestyle{floating}
 {frame=lines,
  float=hptb,
  captionpos=b,
  abovecaptionskip=-0pt}

\lstdefinestyle{py}
 {language=Python,
  showstringspaces=false,
  keywordstyle=\ttfamily\bfseries,
  tabsize=2,
  style=floating,
  belowskip=-0\baselineskip,
  aboveskip=-0\baselineskip,
  morekeywords={}
}

\lstnewenvironment{python}[1][]
 {\lstset{style=py,#1}}{}  

\newcommand{\spy}[1]{\lstinline[style=py]{#1}}

\usepackage[version=0.96]{pgf}
\usepackage{tikz}
\usetikzlibrary{arrows, shapes, backgrounds}
\usetikzlibrary{decorations.pathreplacing}
\usetikzlibrary{shapes.misc}
\usetikzlibrary{petri}\tikzstyle{place}=[circle,thick,draw=black!75,minimum size=5mm]
\tikzstyle{iplace}=[circle,dashed,thick,draw=black!75,minimum size=5mm]
\tikzstyle{itransition}=[rectangle,draw,thick,fill=black,minimum size=1mm]
\tikzstyle{etransition}=[rectangle,draw,thick,minimum size=1mm]
\tikzstyle{ctransition}=[rectangle,draw,color=black!45,thick,fill=black!45,minimum size=1mm]

\tikzstyle{copn}=
 [node distance=1.3cm, >=stealth', bend angle=45, auto,
  font=\fontsize{8}{8}\selectfont]

\newcommand{\eg}{\emph{e.g.,}\xspace}
\newcommand{\ie}{\emph{i.e.,}\xspace}

\newcommand{\copn}[7][]
  {\ensuremath{\mathopen{<}
      {#2}_{#1},{#3}_{#1},{#4}_{#1},{#5}_{#1},{#6}_{#1},{#7}_{#1}
      \mathclose{>}}}

\WithSuffix\newcommand\copn*[1][]{\copn[#1]{P}{T}{f}{f_\circ}{\rho}{m_0}}

\definecolor{author}{rgb}{.5, .5, .5}
\definecolor{comment}{rgb}{.1, .0, .9}
\definecolor{note}{rgb}{.9, .4, .0}
\definecolor{idea}{rgb}{.1, .7, .0}
\definecolor{missing}{rgb}{.9, .1, .0}
\definecolor{OliveGreen}{rgb}{0,0.6,0.3}

\makeatother

\acrodef{AI}{Artificial Intelligence}
\acrodef{API}{Application Programming Interface}
\acrodef{AOP}{Aspect-oriented Programming}
\acrodef{CAS}{Collective Adaptive Systems}
\acrodef{CFG}{Control Flow Graph}
\acrodef{CPU}{Central Processing Unit}
\acrodef{DSL}{Domain-Specific Language}
\acrodef{DNN}{Deep Neural Networks}
\acrodef{DQN}{Deep Q-learning}
\acrodef{LC}{Large Class}
\acrodef{LOC}{Lines of Code}
\acrodef{LLF}{Long Lambda Function}
\acrodef{LMC}{Long Method Chain}
\acrodef{LPL}{Long Parameter List}
\acrodef{LRE}{Local Relative Entropy}
\acrodef{LTCE}{Long Ternary Conditional Expression}
\acrodef{LM}{Long Method}
\acrodef{LSC}{Long Scope Chaining}
\acrodef{ML}{Machine Learning}
\acrodef{MDP}{Markov Decision Process}
\acrodef{MNC}{Multiply-Nested Container}
\acrodef{OOP}{Object-Oriented Programming}
\acrodef{PCA}{Principal Component Analysis}
\acrodef{PDG}{Program Dependence Graph}
\acrodef{RL}{Reinforcement Learning}
\acrodef{SAT}{boolean SATisfiability problem}

\newcommand{\acResetNonTrivial}
  {\acresetall
   \acused{CPU}
   \acused{API}
   \acused{LAN}
   \acused{SMT}
   \acused{GUI}}

\acResetNonTrivial

\begin{document}

\title{Prevalence of Code Smells in Reinforcement Learning Projects
}

\author{\IEEEauthorblockN{Nicol\'as Cardozo}
\IEEEauthorblockA{%\textit{Systems and Computing Engineering Department}\\
\textit{Universidad de los Andes}\\
Bogot\'a, Colombia \\
n.cardozo@uniandes.edu.co}
\and
\IEEEauthorblockN{Ivana Dusparic}
\IEEEauthorblockA{%\textit{School of Computer Science and Statistics} \\
\textit{Trinity College Dublin}\\
Dublin, Ireland \\
ivana.dusparic@tcd.ie}
\and
\IEEEauthorblockN{Christian Cabrera}
\IEEEauthorblockA{%\textit{ Department of Computer Science and Technology} \\
\textit{University of Cambridge}\\
Cambridge, UK \\
chc79@cam.ac.uk}
}

\maketitle

\begin{abstract}
\ac{RL} is being increasingly used to learn and adapt application behavior in many domains, including 
large-scale and safety critical systems, as for example, autonomous driving. With the 
advent of plug-n-play \ac{RL} libraries, its applicability has further increased, enabling integration of 
\ac{RL} algorithms by users. We note, however, that the majority of such code is not developed 
by \ac{RL} engineers, which as a consequence, may lead to poor 
program quality yielding bugs, suboptimal performance, maintainability, and evolution problems for 
\ac{RL}-based projects. 
In this paper we begin the exploration of this hypothesis, specific to code utilizing \ac{RL}, analyzing 
different projects found in the wild, to assess their quality from a software engineering 
perspective. Our study includes 24 popular \ac{RL}-based Python 
projects, analyzed with standard software engineering metrics.
Our results, aligned with similar analyses for ML code in general, show that popular and 
widely reused \ac{RL} repositories contain many code smells (3.95\% of the 
code base on average), significantly affecting the projects' maintainability. 
The most common code smells detected are long method and long method 
chain, highlighting problems in the definition and interaction of  
agents. Detected code smells suggest problems in responsibility 
separation, and the appropriateness of current abstractions for 
the definition of \ac{RL} algorithms.  
\end{abstract}


% $Id: introduction.tex $
% !TEX root = main.tex

\section{Introduction}
\label{sec:introduction}

\acf{RL} is a \ac{ML} approach in which an agent learns to solve 
a problem by trial and error in interaction with the environment. 
\ac{RL} formalizes the agent’s goal, taking 
into account the set of environment states, and the set of possible 
actions to execute in each state. Agents have an associated reward 
model, so that for each given action taken at a given state, agents 
receive a scalar reward, reflecting the desirability of the outcome (\ie positive reward or a punishment). 
The purpose of the agent is to maximize the total cumulative reward 
received to reach its goal, from any given initial 
state~\cite{sutton98}.

In the last years, thanks to the development of open source libraries 
and methods that allow rapid prototyping, \ac{RL} has expanded as a 
tool in many disciplines, like data science, control theory, finance, 
chemistry, and neuroscience. However the rapid grow of these systems 
has lead to prototypes and projects to have very irregular structures 
as they are basically an heterogeneous set of libraries, components, 
program identifiers, data processing functions and training algorithms 
patched with glue code~\cite{jebnoun20}. These projects normally 
introduce \ac{RL} to improve on the processing power or accuracy of an 
existing solution as an immediate goal, but set aside middle and long 
term objectives (\eg easy debugging, maintenance, extensibility, 
future improvements). This loose structure contributes to code 
complexity and technical debt~\cite{sculley15}, which in turn can 
result in severe quality and performance issues of the entire software 
systems~\cite{seaman12, jebnoun20}.
One of the symptoms of poor implementations in software systems are 
code smells~\cite{chen18}. Code smells are bad patterns in source code 
that may signify a violation of fundamental design 
principles~\cite{suryanarayana14}. Code smells slow down software 
evolution due to code misunderstandings, bring maintenance 
difficulties, and increase the risk of defects~\cite{pdr22}. 
The presence of code smells negatively impacts
software quality and can indicate where to apply beneficial 
refactoring. The study of code smells in \ac{ML} projects~\cite{vanoort21, zhang22} 
highlights potential pitfalls that can be decremental for the 
efficiency of the algorithm or even introduce bugs in the program. 
Identified code smells in \ac{ML} projects include coding notation 
misused, coding omissions, errors in function calls, and problems in data 
processing. While current studies have identified the existence of 
general code smells and \ac{ML}-specific code smells, no evaluation 
dedicated to \ac{RL} exists.

In this paper we carry a preliminary empirical study of the code 
quality of \ac{RL} projects, through the identification of code 
smells. The purpose of this is to evaluate our hypothesis: 
\emph{\ac{RL} projects suffer from poor code quality, which may lead 
to maintenance and evolution problems}. We analyze \ac{RL}-based systems with respect to eight 
software quality metrics~\cite{lanza07} related to the code organization, 
abstraction, and expression of programs (\fref{sec:quality}). The 
calculated metrics serve as a proxy to assess the maintainability and 
evolution of \ac{RL}-based projects. We evaluate 24 GitHub \ac{RL}-based Python projects using 
Q-learning and \ac{DQN}, as the most popular and straight-forward \ac{RL} algorithms 
(\fref{sec:evaluation}). Our results identify a total of 1090 code smells across all projects. The most 
common smells found are related to the definition and access to the entities in \ac{RL} projects. This 
exposes a violation of the single responsibility principle, and bad modularization of the 
projects, which is directly related to problems in the maintainability 
and evolution of the projects.
 
The results obtained from the analysis support our hypothesis that \ac{RL}-based projects suffer 
from poor quality, pointing out to the need for specialized analysis 
tools and programming languages for the development of such projects. We conclude the study 
(\fref{sec:lang}) by proposing potential research avenues for addressing this shortcoming, through, 
\eg dedicated static analysis tools, or specialized programming languages.

% !TEX root = main.tex

\section{Assessing Quality of \ac{RL} Projects}
\label{sec:quality}

The development of \ac{RL} projects is occurring mainly in Python. As 
a consequence, we focus the evaluation of our work on the analysis of 
Python-based \ac{RL} projects. 

Our analysis consists of eight code smells related to code 
organization, abstraction, and expression concerns, which gather information about code modularity 
and maintainability. The first five code smells 
are related to object-oriented programming in general, and the last 
three are specific to Python code smells detection~\cite{chen18}. 

  \paragraph*{\textbf{\ac{LM}}} A method or a function that is overly long. This 
impacts the understandability and testability of the code. Long 
methods tend to violate the single responsibility principle, which 
means they have more than one reason to change, using many 
temporary variables and parameters, which makes them more 
error-prone~\cite{lanza07}.

  \paragraph*{\textbf{\ac{LC}}} A class that is overly long. Similar to \ac{LM}, this 
could mean that the class is violating the single responsibility 
principle. This can have a negative impact on the understandability 
and reusability of the functions that the class 
encapsulates~\cite{fowler19}.
 
\paragraph*{\textbf{\ac{LPL}}} 
A method or a function that has a long parameter 
  list. A long parameter list can be inconsistent, difficult to use and 
  understand, and requires continuous changes as more data is 
  needed~\cite{fowler19}.

\paragraph*{\textbf{\ac{LMC}}} 
An expression that is accessing an object using the 
  dot operator through a long sequence of attributes or 
  methods~\cite{brown98}. The presence of this code smell can reveal an 
  unnecessary coupling between classes.  Changes to the intermediate relationships could 
  force undesirable changes on other classes in the chain. %To illustrate this suppose we 
  %have classes \spy{A}, \spy{B}, \spy{C}, and \spy{D}, in which an 
  %instance of \spy{A} needs data from \spy{D}, accessed through 
  %delegations to \spy{B} and \spy{C}, as follows:
  %\begin{python}[frame=none]
   %   a.getB().getC().getD().getData()
  %\end{python}

\paragraph*{\textbf{\ac{LSC}}}
A method or a function that is 
  multiply-nested~\cite{fowler19}. This hinders function readability 
  and affects testability. Like before Multiply-nested functions tend 
  to violate the single responsibility principle.

\paragraph*{\textbf{\ac{LTCE}}} 
A ternary conditional expression that is overly 
  long. The ternary operator defines a conditional expression in Python 
  in the form: \spy{X if C else Y}
  This is a concise way of writing short and simple conditional 
  statements. However, if the expression is too long it can be harder 
  to read and locate than a normal \spy{if} statement~\cite{chen18}.
  
\paragraph*{\textbf{\ac{MNC}}} 
A container (\eg set, list, tuple, dictionary) that is 
  multiply-nested. This produces expressions for accessing an object 
  with a long chain of indexed elements. This affects readability and 
  is error-prone when traversing multidimensional arrays~\cite{chen18}.
  
\paragraph*{\textbf{\ac{LLF}}} 
A lambda function that is too long. Lambda is a 
  Python expression used for the creation of anonymous functions at 
  runtime. These functions tend to be harder to read and debug than 
  local ones because they do not have reference names, and are 
  contained in one line. The problems of long lambda functions are 
  related to those of \ac{LM}~\cite{pyguide}.

Taking these metrics into account, we use the quality profile shown in 
\fref{tab:thresholds} as taken from 
the literature~\cite{lanza07, chen18}.

\begin{table}[hptb]
  \centering
  \caption{Thresholds for the metrics used}
  %\resizebox{0.8\columnwidth}{!}{%
\begin{tabular}{l | llc}
\toprule
\textbf{Code smell} &
\textbf{Metric} &
\textbf{Threshold} \\
\toprule 
\ac{LM} & Function LOC & 38 \\
\ac{LC} & Class LOC & 29 \\
\ac{LPL} & Num. parameters & 5 \\
\ac{LMC} & Length of message chain & 5 \\
\ac{LSC} & Depth of closure & 3 \\
\ac{LTCE} & Num. characters & 54 \\
\ac{MNC} & Depth of nested container & 3 \\
\ac{LLF} & Num. characters & 48 \\
\bottomrule
\end{tabular}%
%}

  \label{tab:thresholds}
\end{table}

% !TEX root = main.tex

\section{Methodology}

\begin{figure}[htpb]
	\centering
	\includegraphics[width=\columnwidth]{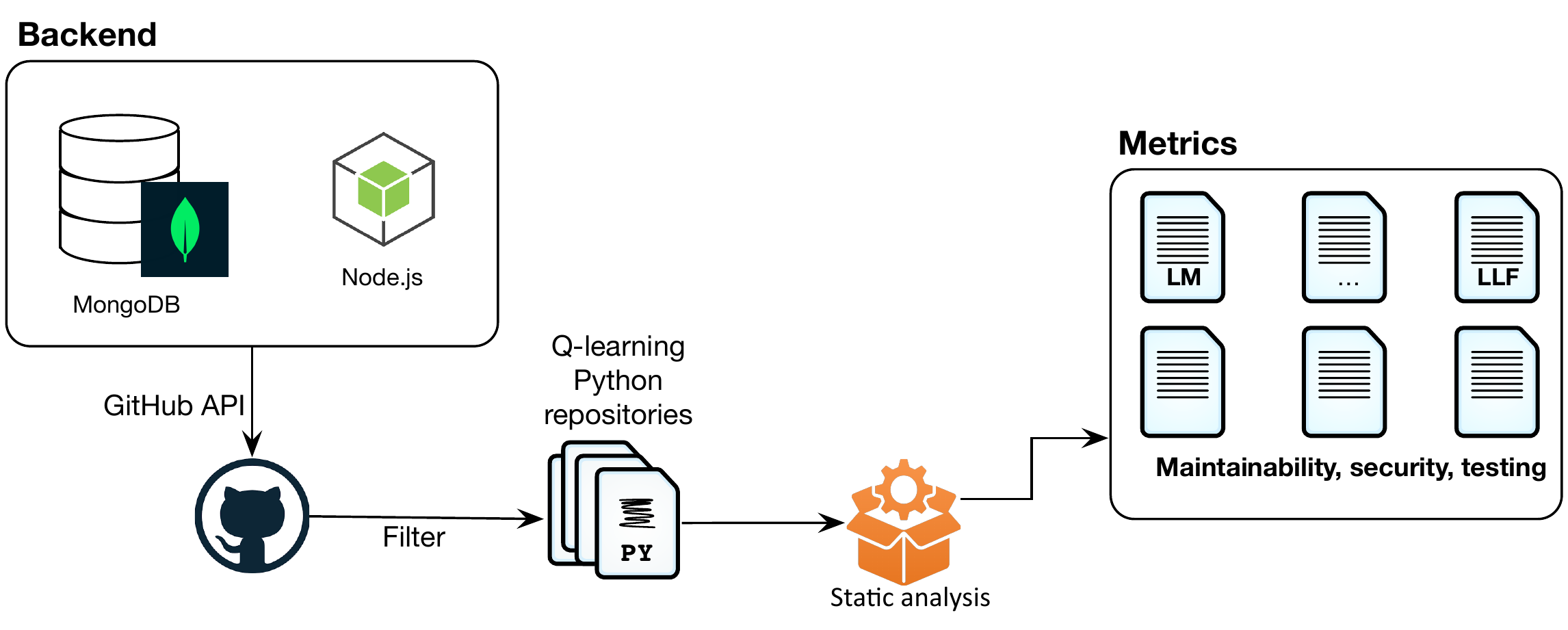}
	\caption{Repository mining, selection, and analysis process}
	\label{fig:extraction}
\end{figure}

The process used to analyze the code quality of the repositories 
is depicted in \fref{fig:extraction}. We evaluated a data set of 24 \ac{RL} Python projects to test our 
hypothesis about their code quality.\footnote{\ac{RL} repositories data set: \url{https://doi.org/10.5281/zenodo.4584135}} 
The data set is split in two subsets to asses the code quality of projects with different characteristics 
in terms of size and developer expertise. Such division allows us to understand if the results 
of the code quality study are attributed to developers or the 
intrinsic complexity of \ac{RL}-based systems. The first 20 projects 
are extracted from open-source GitHub repositories. These correspond to 
the \ac{RL} repositories with the highest number of stars on GitHub 
that are maintained by general GitHub users. To obtain these, we used the GitHub API to filter  
active and popular repositories that implement the Q-learning 
algorithm. We manually filtered the repositories to assure the 
repositories are implemented applications, rather than project 
templates, and do implement the Q-learning algorithm. 
The remaining 4 projects correspond to the ACME 
repositories~\cite{hoffman20}, a research framework developed 
by \ac{RL} engineers,\footnote{\url{https://github.com/deepmind/acme/tree/master/examples}} 
used for the comparison with the \ac{RL} projects found in the wild. We then 
analyze the entire repositories' quality as a whole, using the metrics described in 
\fref{sec:quality}, following the thresholds defined in \fref{tab:thresholds}.

%Additionally we use SonarQube and Prospector on the repositories, to further analyze them with respect to their maintainability, testability, security, and flexibility.

%%
\section{Results and Analysis}
\label{sec:evaluation}

%%%%
\subsection{Results}

The results of our analysis (\fref{tab:qa-results}) show a 
considerable quantity of code smells detected for the 24 \ac{RL} 
projects. The number of code smells identified in the 20 projects 
gathered in the wild significantly differs from the number of code smells identified in the 
ACME projects. This first finding suggests that the projects developed by \ac{RL}
users, rather than by \ac{RL} engineers or developers of purpose could indeed impact negatively 
the quality of \ac{RL}-based systems, supporting our hypothesis.

The results in \fref{tab:qa-results} are ordered from the highest to 
the lowest number of detected code smells. Taking into account the 
first set of repositories, extracted from GitHub, the project with the 
highest number of code smells is Rl\_trading with 300 smells in 35 
files, and the one with the lowest number of code smells is 
Q-trader with 4 code smells in 4 files. This confirms that it is more 
likely for projects with a larger code base
to exhibit more code smells. However, on average, the percentage of 
code smells in the projects is $4.16\%$ (with standard deviation 
$\sigma^2=4.52\%$) of the code base. In proportion, the projects with 
the largest amount of code smells, relative to the code base size, is 
Deep-q-rl in which $22.63\%$ of the code base contains code smells.  
The results obtained from the ACME projects show the 
project with the highest number of code smells is Offline Agents with 
16 smells in 6 files. The project with the lowest number of code 
smells is Continuous control with 0 smells in 7 files. 
The percentage of code smells in the projects is on average $2.90\%$ 
(with standard deviation $\sigma^2=3.02\%$) of the code base. The 
the largest amount of code smells, relative to the code base size, is 
found on Continuous control where $7.14\%$ of the code base 
contains code smells.

We observe that the repositories extracted from GitHub contain 
significantly more code smells than those in the ACME framework. This 
suggest that the code generated by \ac{RL} engineers indeed has a better 
quality than that of the other repositories. Nonetheless, the 
proportion of code smells with respect to the code base is similar. 
This suggest that there is an intrinsic complexity in \ac{RL} 
algorithms that is currently not being addressed.

When we analyze the most recurrent code smells, we observe a similar 
trend between the data sets. The four most recurrent code smells detected 
in the extracted GitHub repositories are \acf{LM} 
(detected 288 times), \acf{LC} (217), \acf{MNC} 
(176), and \acf{LPL} (145).
Similarly, The four most common code smells for the Acme examples are 
\acf{MNC} (detected 8 times), \acf{LLF} (5), \acf{LM} 
(4), and \acf{LPL} (2).

The shared recurrent code smells between the extracted GitHub projects 
and the ACME framework projects again shades lights to an 
intrinsic complexity in the definition and interaction between 
entities of \ac{RL} algorithms that cannot be expressed at the 
appropriate abstraction level at the moment.

\begin{table*}[hptb]
  \centering
  \caption{Results from the quality analysis of the collected repositories}
  \resizebox{\linewidth}{!}{%
\begin{tabular}{l | cc | cccccccc | c  }
\toprule
\textbf{Project} &
\textbf{Num. files} & 
\textbf{LOCs} &
\textbf{\ac{LM}} &
\textbf{\ac{LC}} &
\textbf{\ac{LPL}} &
\textbf{\ac{LMC}} &
\textbf{\ac{LSC}} &
\textbf{\ac{LTCE}} &
\textbf{\ac{MNC}} &
\textbf{\ac{LLF}} &
\textbf{Total code smells} \\

\toprule

\textbf{GitHub projects} &  \\
Rl\_training & 34 & 8172 & 102 & 80 & 57 & 8 & 36 & 0 & 13 & 4 & 300  \\
Deer & 42 & 4453 & 35 & 43 & 22 & 38 & 8 & 5 & 60 & 0 & 211  \\
Dissecting-reinforcement-learning & 45 & 4101 & 39 & 15 & 12 & 6 & 17 & 0 & 17 & 1 & 107 \\
MyDeepLearning & 34 & 2773 & 41 & 8 & 13 & 4 & 6 & 0 & 18 & 2 & 92 \\
Arnold & 24 & 2516 & 29 & 15 & 6 & 4 & 8 & 12 & 15 & 0 & 89\\ 
QA-deep-learning & 15 & 2077 & 8 & 8 & 9 & 17 & 10 & 1 & 13 & 1 & 67\\
Deep-q-rl & 41 & 243 & 12 & 8 & 10 & 3 & 1 & 0 & 21 & 0 & 55\\
Stock\_market\_reinforcement\_learning & 5 & 454 & 6 & 5 & 1 & 1 & 2 & 14 & 3 & 0 & 32\\
Self-Driving-Car-AI & 3 & 453 & 2 & 7 & 0 & 0 & 0 & 0 & 7 & 1 & 17\\
Pytorch-dqn & 9 & 489 & 2 & 5 & 1 & 0 & 2 & 2 & 2 & 1 & 15\\
Snake-ga & 2 & 342 & 2 & 3 & 3 & 0 & 1 & 0 & 3 & 0 & 12\\
Deep-q-learning & 3 & 243 & 0 & 3 & 3 & 0 & 5 & 0 & 0 & 0 & 11\\
Qlearning4k & 10 & 479 & 1 & 5 & 4 & 0 & 0 & 0 & 1 & 0 & 11\\
DRL-FlappyBird & 5 & 485 & 3 & 3 & 0 & 3 & 0 & 0 & 1 & 0 & 10\\
Tetris-deep-Q-learning-pytorch & 4 & 415 & 2 & 1 & 0 & 0 & 4 & 1 & 0 & 1 & 9\\
Async-rl & 5 & 475 & 2 & 1 & 2 & 0 & 2 & 0 & 1 & 0 & 8\\
Flappy-bird-deep-Q-learning-pytorch & 5 & 304 & 2 & 1 & 0 & 0 & 0 & 2 & 1 & 1 & 7 \\
Gym-anytrading & 8 & 321 & 0 & 3 & 0 & 0 & 2 & 1 & 0 & 0 & 6\\
Q-Learning-for-Trading & 5 & 211 & 0 & 2 & 2 & 0 & 1 & 0 & 0 & 0 & 5\\
Q-trader & 4 & 143 & 0 & 1 & 0 & 0 & 0 & 3 & 0 & 0 & 4\\
%Awesome-monte-carlo-tree-search & 0 & 0 & 0 & 0 & 0 & 0 & 0 & 0 & 0 & 1 & 4 \\
\midrule
Total & 303 & 29149 & 288 & 217 & 145 & 84 & 105 & 41 & 176 & 12 & 1068\\
\toprule
\textbf{ACME projects} & \\
Offline Agents & 9 & 711 & 4 & 0 & 2 & 0 & 0 & 0 & 5 & 5 & 16\\
Behaviour Suite & 3 & 137 & 0 & 0 & 0 & 0 & 0 & 0 & 3 & 0 & 3\\
Open Spiel & 1 & 42 & 1 & 0 & 0 & 1 & 0 & 1 & 0 & 0 & 3\\
Continuous control & 7 & 413 & 0 & 0 & 0 & 0 & 0 & 0 & 0 & 0 & 0\\
%Discrete Agents (Atari) & 0 & 0 & 0 & 0 & 0 & 0 & 0 & 0 & 0 & 3 &  & 0 \\
%Gym & 0 & 0 & 0 & 0 & 0 & 0 & 0 & 0 & 0 & 3 &  & 0 \\
\midrule
Total & 20 & 1303 & 5 & 0 & 2 & 1 & 0 & 1 & 8 & 5 & 22\\
\bottomrule
\end{tabular}%
}

  \label{tab:qa-results}
\end{table*}

\subsection{Analysis}

The ACME projects have a lower number of detected 
code smells (\ie 0.8 code smells per file) than the projects extracted 
from GitHub (\ie 3.15 code smells per file). This may be attributed to 
the fact that ACME projects are implemented by \ac{RL} engineers with the 
purpose of being reused by other developers, and therefore have a better 
code quality standard.

Nonetheless, three of the four most recurrent code smells (\ie 
\ac{MNC}, \ac{LM}, and \ac{LPL}) are shared between the two sets of 
projects. The nature of these code smells lies in how programming 
entities access and share information between them. The other common 
code smells \ac{LC} and \ac{LLF} are related to the definition of the 
program entities and their behavior. This finding suggests that there 
is an intrinsic complexity of the systems which is not being captured 
by the code. We observe that the single responsibility principle is 
not being respected according to the number of detected code smells of 
the type \ac{LPL}, \ac{LM}, \ac{LC}, \ac{LLF}. Additionally this 
suggest a low cohesion and a large coupling between program entities. 
The high occurrence of the \ac{MNC} code smell suggests that the 
representation of the state and actions are taking place in 
high-dimensionality structures that are difficult to understand, 
manipulate, and modify. For example, the code of the Deer project to 
accesses the agent's state (\fref{lst:python}) is cumbersome and 
difficult to understand (\eg a \ac{MNC} code smell).

\begin{python}[label={lst:python},
  caption={Deer project array traversing code example}]
  [(player.x_change == 0 and player.y_change
      == -20 and ((list(map(add,player.
      position[-1],[20, 0])) in player.
      position) or
  player.position[ -1][0] + 20 > (game.
      game_width-20)))
\end{python}
      
\ac{MNC} is one of the code smells appearing in most of the 
projects. If we take into account that the ACME projects are more 
concise and focused on the \ac{RL} use cases and applications, the 
fact that \ac{MNC} is the most frequent code smell on those examples, 
suggests that its occurrences are linked to the basic structure of 
\ac{RL} applications. This is reinforced by the fact that the files 
that tend to have this code smell are related to the definition and 
uses of environments, states, rewards, and transition vectors used in 
\ac{RL}. This can be observed for instance in the Deer project, which 
has the highest number of \ac{MNC}. \fref{tab:mnc} shows the \ac{MNC} 
code smells for each file of the Deer project. As can be seen, 29 of 
the 60 \ac{MNC} detected, are part of the environment files.

\begin{table}[hptb]
  \centering
  \caption{\ac{MNC} for individual files in the Deer project}
  %\resizebox{\columnwidth}{!}{%
\begin{tabular}{lc}
\toprule
\textbf{File} &
\textbf{\ac{MNC}} \\
\toprule 
/examples/test\_CRAR/simple\_maze\_env.py & 17 \\
/deer/default\_parser.py & 9 \\
/examples/test\_CRAR/catcher\_env.py & 8 \\
/examples/test\_CRAR/run\_catcher.py & 4 \\
/examples/maze/run\_maze.py & 4 \\
/examples/MG\_two\_storages/run\_MG\_two\_storages.py & 4 \\
examples/test\_CRAR/run\_simple\_maze.py & 2 \\
/examples/ALE/run\_ALE.py & 2 \\
/deer/learning\_algos/CRAR\_keras.py & 2 \\
/examples/toy\_env/run\_toy\_env.py & 1 \\
/examples/ALE/ALE\_env\_gym.py & 1 \\
/examples/ALE/ALE\_env.py & 1 \\
/examples/maze/maze\_env.py & 1 \\
/examples/gym/run\_mountain\_car.py & 1 \\
/examples/gym/run\_mountain\_car\_continuous.py & 1 \\
/deer/agent.py & 1 \\ 
/examples/gym/run\_pendulum.py & 1 \\
\midrule
Total MNC from environment files only & 29 \\
Total MNC from all files & 60 \\
\bottomrule
\end{tabular}%
%}

  \label{tab:mnc}
\end{table}

\subsection{Threats to Validity}

As a preliminary study, a threat to external validity is on the 
generalization of our results from two perspectives. The selection of 
the repositories focuses on the popularity of the repositories, but 
other criteria, as forks or activity may yield a larger set of 
repositories with different characteristics. A similar conclusion can 
be reached from the ACME examples, which are just a small set of 
professional projects.
Another generalization bias may be on the type of projects selected. We keep the evaluated 
repositories to the use of Q-learning. Other \ac{RL} implementations 
(\eg SARSA) may also present different characteristics with respect to 
the code structure and identified smells.

The code smells detected, are the result of using experience-based 
thresholds defined by Python developers~\cite{chen18}. However, these 
thresholds could not be appropriate for the specific case of  
\ac{RL} projects due to the particular features of \ac{RL} itself, 
representing a construct validity threat.

% !TEX root = main.tex

\section{Road Map for Higher Quality \ac{RL}}
\label{sec:lang}

The results show a high presence of code smells hindering the quality 
of \ac{RL} projects. We identify two main research avenues to improve 
the quality of \ac{RL} projects. 

\paragraph*{\textbf{Static Analysis Tools}}
First, we recognize that it is necessary to create specific metrics 
and thresholds for \ac{RL} quality analysis. As observed from our 
analysis, the recurrent code smells persist in the two data sets. One 
possible reason for this commonality may be the complexity of \ac{RL} 
algorithms, which have different characteristics than OO 
algorithms. Existing metrics and thresholds may not correctly capture 
the characteristics of \ac{RL}. Therefore, specialized metrics and 
thresholds can offer a more faithful view of \ac{RL} projects quality. 
Additionally, the metrics should come with the associated  
static analysis tools. A full study of software engineering best 
practices with a catalog of \ac{RL}-specific metrics and thresholds is 
required.

\paragraph*{\textbf{Programming Languages}} 
We observed two characteristics of \ac{RL} projects that influence the presence of code smells. 
\begin{enumerate*}[label=(\arabic*)]
\item Definitions of functions and classes 
are often long in \ac{RL} projects. This affects the maintainability 
and testability of the projects. 
\item The current state of developing \ac{RL} systems relies in basic 
data structures abused to represent the many dimensions required 
to express the environment, actions, and goal of an agent.  
\end{enumerate*}
The definition of a specialized programming language can offer more 
appropriate abstractions for the interaction between the agents and 
the environment can make definitions more concrete.   
Additionally, new programming abstractions can open the possibility to 
capture the ever-changing conditions between the agents and the 
environment. A declarative definition of actions and rewards (\eg by means of a 
predicate or lambda function, to dictate new behavior) could be more easily 
modified at run time to open up the system flexibility. 
For example, such flexibility could open the possibility for lifelong learning~\cite{khetarpal22}, 
adaptation evolution, and agent's goal adaptation~\cite{cardozo21}.

% $Id: related.tex 
% !TEX root = main.tex

\section{Related Work}
\label{sec:related}

%ivana notes: i'd move this section before - so can pick up on it in conclusions of our work, that it could just be lack of training

%\subsection{Quality Analysis of \ac{ML} Projects}

The increasing interest in \ac{ML} generated interest to apply software 
engineering to it~\cite{menzies19}, giving rise to different quality 
analysis proposals, which we discuss in the following.

A general empirical study on the quality requirements from \ac{AI} 
practitioners~\cite{golendukhina22} highlights the 12 most pressing 
issues identified from interviews conducted with industry \ac{AI} 
practitioners. This study is motivated by the identification the lack 
of proper training, practices, and processes in \ac{AI} based projects. 
From the 12 categories identified, three of them are related to code 
quality (\ie non-deterministic Python behavior, independencies between 
projects/modules, and lack of clear code). This reinforces our 
hypothesis that code quality is problematic for developing \ac{RL} 
systems.

The analysis of code smells for general \ac{ML} projects highlights 20 
code smells recurrent in \ac{ML} projects as extracted from a standard 
Python linter~\cite{vanoort21}. Additionally, they highlight 22 
specific code smells mined from the literature, bug databases, forums, 
and blogs for \ac{ML} applications~\cite{zhang22}. The \ac{LM}, 
\ac{LPL}, \ac{LMC}, and \ac{LLF} code smells all feature in the 
identified lists, suggesting a common complexity for accessing and 
expressing both \ac{RL} and \ac{ML} algorithms.

Assuring quality has focused on testing \ac{AI} 
systems~\cite{murphy06}. Different proposals exist for bug detection, 
identification, and fixing in \ac{ML} projects. There are 7 major bug 
types identified for \ac{ML} projects, extracted from the empirical 
analysis of 3 \ac{ML} projects~\cite{sun17}.  
Four of the bug types are directly related to the code quality (\ie 
variable bugs, design defects, performance optimization, and memory 
overflow). A similar study analyzes TensorFlow-related bugs in deep 
learning applications~\cite{zhang18}, with 175 bugs related to code 
quality detected. Similarly, common bug fixing patterns in \ac{DNN}
~\cite{islam20} include errors in the dimensionality of data or the 
connectivity of the layers in the network, which lead to  
application crashes. Such bugs, are directly associated with the 
expressiveness of the tools to develop \ac{DNN} projects.

Closely related to our study,~\citet{wang20} analyze the  quality of 
Jupyter Notebooks code, identifying issues in their design and 
maintainability. As a conclusion, there is a 
need for analysis tools to ensure the quality of Jupyter notebooks 
code. However, another argument over the results could be to improve 
the expressiveness of Jupyter notebooks so that there is no accidental 
complexity when structuring \ac{ML} programs. 

%%

%ivana notes: this probably doesnt need to be submission - maybe paragraphs, since they're so small subsections

%\subsection{Programming Languages for \ac{ML}}

%Identifying the problems raised by code level bugs and lack of quality 
%of \ac{AI}/\ac{ML} projects, highlights the need for new abstractions 
%and tools assuring code quality during the development 
%process~\cite{golendukhina22}. A possibility to address this issues is 
%tackle them from the programming language perspective.

There is a study on the impact of programming languages 
to \ac{ML} projects~\cite{sztwiertnia21}, identifying that programming 
languages can add to the bugs detected for \ac{ML} 
projects. The results show that 15\% of code quality bugs in average
can be attributed to the programming language. This suggests, that 
the use of more appropriate programming languages can help reduce the 
bugs in \ac{ML} projects, hence improving their quality.

Our preliminary code quality analysis of \ac{RL} projects, is inline 
with existing analyses for general \ac{ML}. We reach a similar 
conclusion that code smells hinder the quality 
of \ac{RL} projects. This highlights the need for code analysis tools 
specific to mange the characteristics of \ac{RL} projects, as well as 
the need for better programming abstractions.

% $Id: conclusion.tex 
% !TEX root = main.tex

%%
\section{Conclusion and Future Work}
\label{sec:conclusion}

This paper presents a code quality analysis of 24 \ac{RL} GitHub 
projects, extracted from popular repositories (20 top \ac{RL} GitHub 
repositories), and developed for the purpose of building an \ac{RL} framework (4 ACME 
examples). We analyze the projects using eight software metrics related 
to the definition, access, and interaction of program entities, as a 
proxy of code quality.

The results show that:
the recurrent code smells for the top 20 \ac{RL} GitHub projects are 
\acl{LM}, \acl{LC}, \acl{MNC}, and \acl{LPL}. The recurrent code 
smells for the ACME examples are \acl{MNC}, \acl{LLF}, \acl{LM}, and 
\acl{LPL}.
We observe that the projects developed for the ACME framework
examples present significantly less code smells than the projects 
developed by \ac{RL} users. This supports our hypothesis that 
\ac{RL} users produce less-quality code, which is more prone 
to errors, and potentially present higher maintainability, 
reusability, and extensibility costs. 
Nonetheless, when looking at the recurrent code smells across all 
repositories, we observe a commonality over code smells related to the 
definition and use of program entities and behavior. This suggests, 
that there is an intrinsic complexity in developing \ac{RL} algorithms that should be addressed.
For example, the presence of \acl{LM} and \acl{LC}, indicates that 
separating the responsibilities of program entities is a difficult 
task. This makes classes' and methods' definition complex, leading to 
code that is harder to understand, reuse, and maintain.
The \acl{MNC} and \acl{LPL} code smells indicate that the basic 
structures used to define \ac{RL} systems (\ie environments, states, 
rewards, and transition vectors) are not appropriate to capture the 
dimensionality of the problem at hand, making \ac{RL} projects hard to 
understand and maintain.

As a future work, the research roadmap laid from this study suggest:
First, the need for specialized software quality analysis metrics and tools to 
support \ac{RL} development. Second, the expression of \ac{RL} problems 
and algorithms is intrinsically complex for the current available 
development tools (\ie programming languages). Therefor the creation of 
more expressive abstractions is required to lower such complexity, and 
increase the maintainability of \ac{RL} projects.

\section*{Acknowledgment}
We would like to thank Rafael Bermudez and Andres Felipe Carrion for their contribution to this work.

\printbibliography

\end{document}